\documentstyle[11pt,newpasp,twoside,epsf]{article}
\def\edcomment#1{\iffalse\marginpar{\raggedright\sl#1\/}\else\relax\fi}
\reversemarginpar

\begin{document}
\title{In Search of Evidence for Toroidal B Fields Associated
with the Jets of AGN}
\author{Denise C. Gabuzda and \'Eamonn Murray}
\affil{Physics Department, University College Cork, Cork, Ireland}

\begin{abstract}
Evidence is mounting that many of the transverse jet {\bf B} fields
observed in BL~Lac objects on parsec scales represent the dominant
toroidal component of the intrinsic jet {\bf B} fields. If this is the
case, this may give rise to rotation-measure (RM) gradients across the
jets, due to the systematic change in the line-of-sight component of
the jet {\bf B} field. We have found evidence for such RM gradients in
several BL Lac objects. We discuss these new results, together with 
some of their implications for our understanding of the pc-scale jets 
of AGN.
\end{abstract}

\section{Introduction}

BL~Lac objects are a subset of Active Galactic Nuclei that are observationally
similar to radio-loud quasars in many respects, but display systematically
weaker optical line emission.  BL~Lac objects are also characterised
by strong and variable polarisation at ultraviolet through radio wavelengths.
The radio emission and much of the higher-frequency emission is almost
certainly synchrotron radiation.

VLBI polarization observations of radio-loud BL~Lac objects have shown
a tendency for the dominant magnetic ({\bf B}) fields in the parsec-scale
jets to be transverse to the local jet direction (Gabuzda, Pushkarev, \&
Cawthorne 2000 \& references therein). This has often been interpreted as
evidence for relativistic shocks that enhance the {\bf B}-field component
in the plane of compression, perpendicular to the direction of propagation
of the shock (Laing 1980; Hughes, Aller, \& Aller 1989).

It has been suggested more recently that the transverse jet {\bf B}
fields of BL~Lac objects often correspond to the toroidal {\bf B}-field
component of the jet itself (e.g., Gabuzda 1999, 2003; Gabuzda \& Pushkarev
2002). Such fields would come about naturally, for
example, as a result of ``winding up'' of an initial ``seed'' field
with a significant longitudinal component by the rotation of the central
accreting object (e.g.  Ustyugova et al. 2000; Nakamura, Uchida, \& Hirose
2001).

It is therefore of interest to identify robust
observational tests that can distinguish between transverse {\bf B}
fields due to a toroidal field component and due to shock compression.
One possibility is to search for rotation-measure (RM) gradients {\em across}
the jets, which should arise in the case of a toroidal {\bf B}-field
structure due to the systematic change in the line-of-sight magnetic
field across the jet.  Asada et al. (2002) claim to have detected such
a gradient across the VLBI jet of 3C273.  We present and discuss here 
evidence for transverse RM gradients in a number of BL~Lac objects.

\section{Observations}

We are engaged in an ongoing study of the 34 sources in the complete sample
of northern BL Lacertae objects defined by K\"uhr \& Schmidt (1990).  As a
first step in a search for transverse RM gradients in these
sources, we initially concentrated on several sources that seemed to be
good candidates for such studies, because their jets were 
rich in intensity and polarization structure and relatively well resolved 
in the transverse direction. 

The observations considered here were carried out in February 1997 (0820+225,
1652+398 (Mrk501), 1749+701), April 1997 (1219+285, 1308+326, 1803+784),
and May 2001 (1749+701, 1823+568) at 6, 4 \& 2~cm using the NRAO Very Long 
Baseline Array. The total intensity ($I$) and linear polarization ($P$) 
calibration and imaging were done in AIPS using standard techniques.
We then made 2, 4 and 6-cm matched-resolution (corresponding to the 6-cm 
beam) images of the Stokes parameters $I$, $Q$ and $U$ for each of the 
sources. The $Q$ and $U$ images were combined to make maps of the polarized 
flux and polarization position angle, $\chi$. After subtracting the rotation 
corresponding to the known integrated (assumed to be largely Galactic) 
rotation measures (Pushkarev 1999) at each wavelength, the $\chi$ maps 
at each of the three wavelengths were then used to derive the RM distributions 
using the AIPS task RM. 

\section{Results}

We included 0820+225 in this initial study because its curved jet is well
resolved and very rich in $I$ and $P$ structure, extending tens of 
milliarcseconds from the core (Gabuzda, Pushkarev, \& Garnich 2001). 
Although a comparison of the 4-cm and 6-cm polarization angles indicates 
the presence of a transverse RM gradient beyond the westward bend of the 
jet (Fig.~1), we did not attempt to obtain a three-wavelength RM map 
for this source, since the 2-cm emission was weak beyond the innermost 
jet.

We found reasonable to good evidence for transverse RM gradients in all 
the remaining sources except for 1308+326. Fig.~2 shows our RM map of 
1652+398 as an example, together with plots of the observed 
polarization position angles $\chi$ as functions of the square of the 
observing wavelength, $\lambda^2$, for the two opposite sides and the
center of the VLBI 
jet. These plots show that the $\chi$ values display the linear dependence 
on $\lambda^2$ expected for Faraday rotation. 

\begin{figure}
\plotfiddle{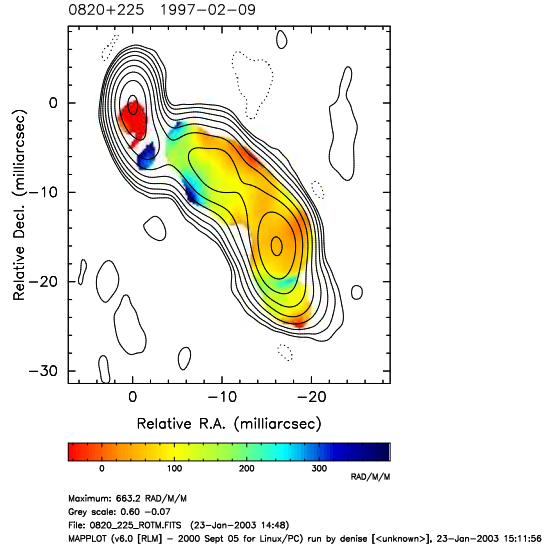}{6cm}{-90}{40}{40}{-180}{220}
\caption{6-cm total intensity contours of 0820+225 superposed on the
two-wavelength rotation-measure distribution obtained by comparing
the polarization angles at 4~cm and 6~cm. }
\end{figure}

\begin{figure}
\plotfiddle{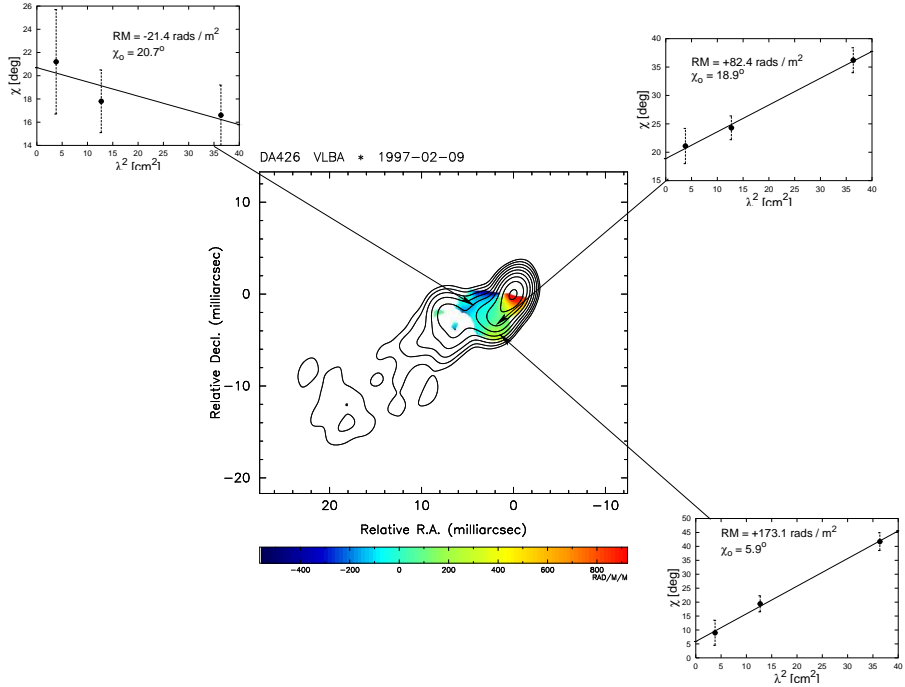}{8cm}{-90}{48}{48}{-190}{265}
\caption{6-cm total intensity contours of 1652+398 superposed on the
rotation-measure distribution derived using our 2~cm, 4~cm and 6~cm 
polarization data. Also shown are plots of the polarization
angle versus $\lambda^2$ for opposite sides and the center of the 
transverse RM gradient.}
\end{figure}

\section{Discussion}

Of the seven sources investigated, we found clear systematic RM gradients 
transverse to the VLBI jets in 1652+398 and 1823+568, with $\lambda^2$
laws being obeyed well in both cases. Some indications of transverse RM
gradients were also found in 1219+285 and 1749+701. 
A comparison of the 4~cm and 6~cm polarization-angle 
distributions for 0820+225 shows a clear transverse gradient in the RM 
distribution, but we were not able to verify this on the basis of data 
obtained at more than two wavelengths. Finally, although our own 
RM map of 1803+784 did not show convincing evidence for transverse RM
gradients, the RM map of this source recently presented by Zavala \& Taylor 
(2003), based on six wavelengths in the VLBA 2~cm and 4~cm bands, clearly 
shows a transverse RM gradient about 4~mas from the VLBI core. Thus, 
we consider this initial search for transverse RM gradients to have been 
successful. 

It is natural to interpret the observed transverse RM gradients as 
reflecting the presence of a toroidal or helical {\bf B} field associated
with these VLBI jets. In this case, the origin of the gradients is the 
systematic change in the line-of-sight {\bf B} field component across the 
jet. The fact that we detected transverse RM gradients
in several of the sources considered in this initial study suggests that
they may be common in BL~Lac objects. 

A helical {\bf B}-field structure
could come about in a natural way as a consequence of ``winding up'' of
a seed field threading the central accretion disk by the joint action of 
the rotation of the accretion disk and the jet outflow (e.g., Ustyugova 
et al. 2000; Nakamura, Uchida, \& Hirose 2001). It is intriguing that
a {\bf B} field with a predominant toroidal component would also come
about if a non-zero current flows along the jet (e.g., Istomin \& Pariev
1996). Indeed, one can turn the problem around, and assert that the 
presence of a substantial toroidal {\bf B}-field component requires that 
there be a non-zero current in the jet!

In the simplest case when we are viewing a toroidal or helical {\bf B} field
``from the side'' (i.e., at $90^{\circ}$ to the jet axis) in the rest
frame of the source and the distribution of thermal electrons is
approximately uniform, we would expect to observe a rotation measure
close to zero along the jet axis (since the line-of-sight {\bf B}-field
component there is close to zero) and RMs of opposite sign on either side
of the axis. This is roughly the behavior shown by the RM distribution
in Fig.~2, where the RM is positive on the southwestern side of the
jet and negative on the northeastern side of the jet. When the {\bf B} 
field is viewed at some other angle to the jet axis, there will still
be a systematic gradient in the RM across the jet, however the gradient
``peak'' will be shifted, and the RM will not necessarily pass through zero.

Note that viewing
the jet at $90^{\circ}$ to its axis in the rest frame of the {\em jet}
is equivalent to viewing the jet at an angle of $\simeq 1/\gamma$ in
the rest frame of the {\em observer}, where $\gamma$ is the Lorentz factor
of the bulk motion of the jet flow. Thus, since we know the jets of AGN 
such as BL Lac objects make relatively small angles to the line of sight,
of order $\theta\simeq 1/\gamma$, it may be that we should expect to see such
relatively symmetrical transverse RM distributions in these sources fairly
often. 

\section{Conclusions}

The results of our initial search for rotation-measure gradients transverse
to the VLBI jets of BL Lac objects have yielded good evidence for such
gradients in several sources. This lends further support to earlier arguments
that the ``transverse'' {\bf B} fields that characterize the jets of these 
objects are associated with toroidal or helical structure of the intrinsic
jet {\bf B} fields. This underlines the view of these jets as fundamentally
electromagnetic structures, and suggests that they may well carry non-zero 
currents. 

The occurrence of Faraday rotation requires both a non-zero line-of-sight 
{\bf B}-field
component and the presence of free electrons in the medium through which
the linearly polarized radiation propagates. This complicates searches
for systematic behavior of the RM distribution across jet structures,
since the density of free electrons is likely to drop off 
with distance from the core (e.g., Taylor 1998, 2000; Reynolds, Cawthorne,
\& Gabuzda 2001; Gabuzda \&
Chernetskii 2003). Higher resolution can be attained by moving toward
observations at shorter wavelengths, but such observations will be less
sensitive to Faraday rotation, since the rotation measure increases in
proportion to $\lambda^2$. It appears that the use of multiple wavelengths
within the VLBA 2~cm, 4~cm and 6~cm bands represents the best available
approach for such studies in terms of providing both good resolution and
good sensitivity to Faraday rotation. We have recently acquired such data
for 0820+225, 1219+285, 1652+398, 1749+701, and 1803+784 in order to 
confirm the presence of transverse RM gradients across their VLBI jets and
to study the properties of these gradients in more detail.

\acknowledgements

We would like to thank Patrick Cronin for help in the preparation of
the figures presented here.

\end{document}